\documentstyle[sprocl,epsfig]{article}
\newcommand{\beq}{\begin{equation}}
\newcommand{\eeq}{\end{equation}}
\newcommand{\beqn}{\begin{eqnarray}}
\newcommand{\eeqn}{\end{eqnarray}}
\newcommand{\bea}[1]{\beq\begin{array}{#1}}
\newcommand{\eea}{\end{array}\eeq}

\newcommand{\be}{\begin{equation}}
\newcommand{\ee}{\end{equation}}
\newcommand{\La}{\Lambda_{QCD}}

\begin{document}
\title{GLUON CONDENSATE AND  BEYOND
\\~\\The 1999 Sakurai Prize
Lecture\,\footnote{Talk at the 1999 Centennial Meeting of the American Physical
Society, March 20-26, on the occasion of receiving the 1999 Sakurai Prize for
Theoretical Particle Physics.}}
\author{V.I. ZAKHAROV}
\address{Max-Planck Institut f\"ur Physik,\\
F\"ohringer Ring 6, 80805 M\"unchen, Germany\\
 E-mail: xxz@mppmu.mpg.de}

\maketitle

\abstracts{
We review briefly and in retrospect the development which brought about
the QCD sum rules based on introduction of the gluon condensate
(M.A. Shifman, A.I. Vainshtein, and V.I. Zakharov (1978)).
}

\section*{ Introduction. QCD '76-'77}

This review is based on the Sakurai prize lecture
given at the Centennial Meeting
of the American Physical Society (Atlanta, March 1999).
The lecture was the second in the series
of three talks related to the Sakurai prize 1999,
and it followed Arkady Vainshtein's summary of the discovery of the
penguins \cite{arkady}.
My main task was to review the QCD sum rules \cite{svz}
within the context of the time when they were uncovered,
that is the years 1976-79. Roughly speaking, the QCD sum rules relate
properties of resonances, such as mass and leptonic width of,
say, $\rho$-meson to the vacuum properties which are parameterized 
in terms of quark and gluon condensates.

By the year 1976, QCD was of course
highly appreciated and quite far developed. One may say that
the basic ideas constituting the present-day understanding
of QCD were already put on the table
although in somewhat less coherent way than we know them today.
In particular, I would emphasize three points related, to
different extent, to what we are going to discuss later:

({\it i}) Asymptotic freedom \cite{af}
was famous by that time. The effective coupling
tends to zero at short distances, or high momenta $Q$:
\be
\alpha_s(Q^2)~\approx~{1\over b_0\ln(Q^2/\La^2)}\label{coupling}
.\ee
Moreover, since up- and down-quarks are practically massless
the only dimensional parameter in QCD is the ultraviolet
cut off $\Lambda_{UV}$.
Hence, it was known very well that the observable hadronic masses
should be generated via the so called dimensional transmutation:
\be
m_N~\approx~(const)\Lambda_{UV}\exp(-b_0/2\alpha_s(\Lambda^2_{UV}))
.\ee
Although the relation looks rather exotic, the QCD sum rules,
as we shall see, make a half way to realize it.

({\it ii}) It was known that the perturbative expansion
by itself, even taken to an infinite order, cannot explain
why the $\eta'$-meson is heavy.
Moreover, an example of a non-perturbative vacuum
field, that is the instanton solution \cite{instanton} has been found.
Generically, instantons do solve the $\eta'$-problem \cite{thooft}.
Also, instability of the QCD vacuum with respect to formation
of nonperturbative color magnetic field had been conjectured 
\cite{matinyan}.

{(\it iii}) The dual-superconductor model of the quark confinement
had been proposed \cite{nambu}. The basic idea behind the model is that
the properties of the QCD vacuum are similar to the properties
of ordinary superconductor. Indeed, if a pair of magnetic charges
is introduced into a superconductor the potential energy
of the pair would grow linearly with the distance $r$ at large distances:
\be
\lim_{r\to \infty}{V(r)}~\approx~\sigma_{\infty}\cdot r\label{lingrowth}
\ee
where $\sigma_{\infty}$ is the tension of the Abrikosov-Nielsen-Olesen string
\cite{ano}. In QCD, a similar phenomenon was postulated to happen,
with a change of magnetic charges to (color) electric, or dual charges.

Because of the asymptotic freedom (AF)
the quantitative predictions of QCD referred to short distances
while access to non-perturbative effects (see, e.g., points
({\it ii}), ({\it iii}) above) was blocked by infrared divergences.
The basic strategy adopted in \cite{svz} was to
exploit the power of perturbative expansions at short
distances and, abandoning for the moment the ambitious
program of tampering and understanding infrared sensitive
contributions, simply parameterize them in terms of a very few numbers.

This mini-review is in three parts:\\
1. Resonance properties and asymptotic freedom.\\
2. Gluon condensate.\\
3. Further developments.

Thus, we are going to
review first sum rules which constrain resonance properties
basing exclusively on the AF \cite{okun}.
Then we will introduce the idea \cite{svz} that it is vacuum condensates,
which limit the validity
of the AF at moderate momenta $Q^2$. Finally, we will highlight
a few topics related to much more recent developments during
last few years. In all the cases we take the freedom of being
subjective and not aiming at completeness of the presentation
to any extent.

\section*{Sum Rules}

It is intrinsic to the method that we are going to use that
one deals not with a particular hadronic state directly
but rather with sum rules. This is because the theoretical
predictions refer to short distances and times where
the effective coupling (\ref{coupling}) is small.
On the other hand, to perform a measurement on a state with a
definite energy one needs long time, because of the uncertainty principle.

The method of sum rules is deeply rooted in
quantum mechanics, and first sum rules are well known since long.
The simplest one seems to be that the probability to find a system in
one of the states is unity.
Similarly, the completeness condition reads as:
\be
|n\rangle\langle n|~=~I
\label{complete}
\ee
where $I$ is the unit operator and $|n\rangle$ is a a complete set of
states.
An example closer to our topic is provided by
the Thomas-Reiche-Kuhn sum rules
for dipole transitions in atoms. One starts with the canonical
commutator
\be
[r_i,p_k]~=~i\delta_{ik}\label{canonical}
\ee
and averages it over a ground S-wave atomic state
$|0\rangle$ with energy $E_0$.
Inserting a complete set of states into
the (see Eq (\ref{complete})) in the left hand side  of
(\ref{canonical}) one immediately arrives at
\be
{3\over m}~=~\Sigma_n(E_n-E_0)|\langle 0|r_i|n\rangle |^2\label{trk}
.\ee
The matrix elements of $r_i$ are measurable in
dipole electromagnetic transitions between the atomic states.

Note that the canonical commutator (\ref{canonical}) is the
same as for free particles. In this sense the situation resembles
QCD where the quarks propagate at short distances the same as free
particles.
However, observing hadrons we would not find much quarks at short distances
since they are predominantly at a characteristic distance of order
$\La^{-1}$. Therefore, to ensure that quarks do not fly away
one has to resort to an external source of quarks such
as electromagnetic current and consider unphysical kinematics
with space-like total momentum of quarks $q$, $-q^2\equiv Q^2\gg \La^2$. Then
according to the uncertainty principle quarks can exist for time
of order
\be
\tau~\sim~{1\over \sqrt{Q^2}}\label{time}
\ee
which is small if $Q^2$ is large.
For consistency, after such time the quarks are to be absorbed by 
another
current, see Fig. 1.
\begin{figure}[h]
 \begin{center}
    \leavevmode
    \epsfig{figure=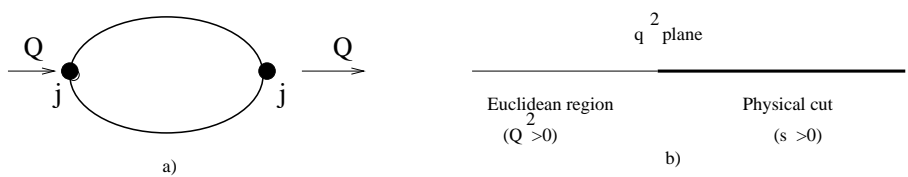,width=12cm}
    \caption{
a) Correlator of currents in the parton model approximation.
     {b) $q^2$ plane.} }
    \label{fig:fig1}
  \end{center}
\end{figure}

In the field theoretical language, we are considering in fact
a correlation function $\Pi_j (Q^2)$:
\be
\Pi_j (Q^2)~=~i\int d^4x~\exp(~iqx)\langle 0|T\{j(x), j(0)\}|0\rangle,
~~~~Q^2\equiv-q^2 \label{correlator}
\ee
where the current $j$ may have various quantum number, like spin, isospin
and for simplicity we do not indicate these quantum numbers, i.e. suppress
the Lorenz indices and so on.

The basic theoretical ingredient is that $\Pi(Q^2)$ at large $Q^2$
can be calculated in the parton model approximation:
\be
\lim_{Q^2\to\infty}\Pi_j(Q^2)~=~\Pi_j(Q^2)_{\rm parton~model}
.\ee
On the other hand, by using
dispersion relations
$\Pi_j (Q^2)$ can be expressed in terms of the absorptive part
which is non-vanishing only for time-like total momentum, $q^2>0$:
\be
\Pi_j(Q^2)~=~{1\over \pi}\int{Im\Pi_j(s)\over s+Q^2}ds\label{dr}
.\ee
The imaginary part is directly observable, provided that
the current $j$ is a physical one. In particular,
in case of the electromagnetic current, $j=j_{el}$ the imaginary
part in Eq (\ref{dr}) is proportional to the
total cross section of $e^{+}e^{-}$-annihilation into hadrons:
\be
Im\Pi_{J_{el}}(s)~=~(const){(\sigma_{tot}(e^{+}e^{-}\rightarrow hadrons)
\over \sigma (e^+e^-\rightarrow\mu^+\mu^-)}\label{st}
.\ee
Upon substitution of (\ref{st}), the Eq. (\ref{dr})
becomes no less a sum rule than, say,
(\ref{trk}). Indeed, $\Pi_{j_{el}}(Q^2)$ is calculable
and the same as for
free particles plus small radiative corrections, while $Im\Pi (s)$
is observable.

It is worth noting that to apply the technique considered
it suffices to ensure that the time which quarks exist
is small indeed. Apart from imposing the condition that
$Q^2$ is large (see Eq (\ref{time})) there exist other possibilities.
In particular, as far as production of heavy quarks is concerned
one can consider $Q^2=0$ since in that case \cite{svz1} $$\tau\sim 1/m_H~.$$
This observation turned in fact crucial for the charmonium sum rules
\cite{okun}. Also,
one might consider a complex value of $q^2$ provided that it is
still far enough from
the cut $s>0$ \cite{poggio}.

After a second thought, one might say, however, that
as far as $Q^2$ is large
the Eq. (\ref{st})
is not so much a sum rule but rather a tautology.
Indeed, Feynman graphs themselves define analytical functions
$\Pi_j (q^2)$. Then evaluating  $\Pi_j$ in deeply Euclidean region
of large $Q^2$ is essentially the same as evaluating the corresponding
cross section within the parton model directly for positive
$s>0$. For example, if we resort to
Fig. 1 both to evaluate the real part of $\Pi (Q^2)$ at large Euclidean
$Q^2$ and the imaginary part at $s>0$ then we get a trivial relation:
\be
\ln{\Lambda^2_{UV}\over Q^2}~=~\int^{\Lambda^2_{UV}}{ds\over s+Q^2}
  .\ee
And, indeed, we do not seem to learn anything new beyond the 
famous parton model
prediction that the total cross section of $e^+e^-$ annihilation
into hadrons is the same as for free quarks. This story can
repeat itself order by order in $\alpha_s(Q^2)$.

Thus, for me personally the whole saga of the sum rules began
not with systematic studies along the lines outlined above but rather
from a conversation with Arkady Vainshtein during one of my
visits to Novosibirsk. For the reasons which I do not remember at all,
we discussed the positronium case. And Arkady was insisting that,
on one hand, we could evaluate the polarization operator to
four-loop order in the Euclidean region (see Fig. 2)
\begin{figure}[h]
 \begin{center}
    \leavevmode
    \epsfig{figure=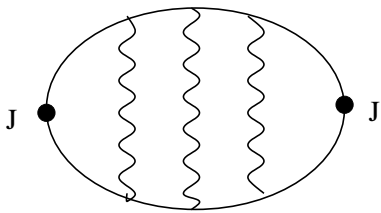,width=5cm}
    \caption{An example of a four-loop graph contributing to a 
correlator of local currents in case of QED. The wavy lines are photons
and the solid lines are electrons (positrons).}
    \label{fig:fig2}
  \end{center}
\end{figure}
while, on the other hand, in the physical cross section we
would have to account for the contributions of the positronium states
which do not arise to any order in perturbative expansion
(see Fig. 3).
\begin{figure}[h]
  \begin{center}
    \leavevmode
     \epsfig{figure=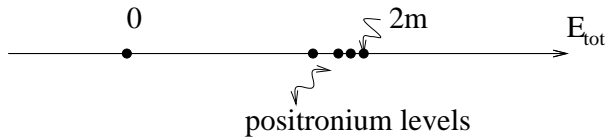,width=8cm}
    \caption{Structure of the $q^2$ plane in case of QED. Apart from the 
cut beginning at $2m_e$, there are positronium states below the threshold
of $e^+e^-$ production.}
    \label{fig:fig3}
  \end{center}
\end{figure}

Indeed, the widths of the positronium states
for transition induced by local currents are proportional to
\be
\Gamma~\sim~|\Psi_{\bar{e}e} (0)|^2~\sim~\alpha_{el}^3,\label{counting}
\ee
where $\Psi_{\bar{e}e}(0)$ is the $\Psi$-function at the origin,
and they do contribute to
the imaginary part of the polarization operator
at the four-loop level.

Looking backwards, it was a proof of non-triviality of the sum rules.
Still, the whole problem seemed to be of pure academic interest at best.
Indeed, we did not expect  at all to learn something new about QED.
Moreover, it was very late at night or, better to say, early in the
morning and, what was most irritating for me, it was a kind of mockery
even to contemplate a possibility
that I would ever be able to evaluate a four-loop graph.

\section*{Asymptotic Freedom vs Resonances}

The positronium story got, however, a dramatic turn sometime later
after discussions at ITEP
when it was realized that the counting of the powers
of $\alpha_s$ looks very different from the counting of
the powers of $\alpha_{el}$ in the positronium case, see
(\ref{counting}). The crucial point is that the effective
coupling (\ref{coupling}) changes rapidly between momenta of
order $Q^2$ and $\La^2$.
The parton model predictions are sensitive to $\alpha_s(Q^2)$.
On the other hand, the resonances are governed by $\alpha_s$
of order unity
at momenta of order $\La$
and not sensitive to the small $\alpha_s(Q^2)$ at all:
\be
\Psi_{\bar{c}c}(0)~\sim~[\alpha_s(Q^2)]^0
 \ee
where $\Psi_{\bar{c}c}(0)$ is the $\Psi$-function at the origin
of the charmonium states.
Thus, even if we evaluate $\Pi_j(Q^2)$ to the lowest order
in perturbation theory, in the dispersive part $Im\Pi_j(s)$
we need to keep the resonance contributions.
Which means in turn
that the asymptotic freedom itself constrains the resonance
properties! Quantum Chromodynamics appeared to be very friendly 
towards people
whose ability to evaluate Feynman graphs does not go beyond one loop.

We were able in fact to include two loops as well, and ended up with sum
rules of the form \cite{okun}:
\be
\int{R_c(s)~ds\over s^{n+1}}\approx~
{A_n\over (4m_c^2)^n}\left(1+B_n\alpha_s(m_c^2)\right),\label{psi}
\ee
where $R_c$ is the contribution of the current of
the charmed quarks into the ratio $R(s)$, $A_n,B_n$ are calculable numbers,
 and $m_c$ is the mass of the charmed quark
(it goes without saying that the only ``heavy'' quark known
at that time was the
$c$-quark).
The integer number $n$ corresponds to the $n$-th derivative from
(\ref{dr}) and we have chosen $Q^2=0$
in case of heavy quarks.
Moreover, as is emphasized above we are to keep the charmonium
contribution in the left-hand side of Eq. (\ref{psi}).
Finally, the $\alpha_s(m_c^2)$ correction corresponds to the 
two-loop contribution, or one-gluon exchange. It was not difficult to 
adapt known QED results to include two loops as well. The idea was to 
control violations of
the AF through the coefficients $B_n$.

The central point about the sum rules (\ref{psi}) 
is that it is not only so that we are allowed to keep the
resonances but that they turn to dominate the sum rules.
Indeed, a single glance at the experimental cross section
of (hidden) charm production in $e^+e^-$ annihilation reveals
that the $J/\psi$ is a huge resonance
by far overshadowing the continuum production
of the $D$-mesons.
Saturating the sum rules by a single resonance we got
relations like \cite{okun}
\be
\Gamma_{ll}(J/\psi)~\approx~{4\alpha_{el}^2\over 27\pi}{(A_3)^4
\over (A_4)^3}M_{j/\psi}~\approx~5~KeV.\label{number}
\ee
where we neglected the $\alpha_s$ corrections altogether.
At a closer look, there were many nontrivial issues involved.
Victor Novikov, Lev Okun, Misha Shifman, Arkady Vainshtein,
Misha Voloshin and myself, we worked enthusiastically
together and summarized the findings in
an issue of the ``Physics Reports'' \cite{okun}.

After the initial euphoria, 
however, we began to evaluate the results more sober.
To begin with, we were not without predecessors in relating
the resonance properties to the bare quark cross section.
The first in the line seemed to be J.J. Sakurai who had said that
the leptonic widths of vector mesons like the $\rho$
correspond to the quark cross section smeared over the interval of
$s$ till the next resonance, like the $\rho'$ \cite{sakurai}.
Indeed, our relations in case of the light quarks
with isotopic spin $I=1$ looked as:
\be
\int ds\exp(-s/M^2)R^{I=1}(s)~
\approx~{3\over 2} M^2\left(1+{\alpha_s(M^2)\over \pi}\right)\label{rho}
\ee
where $M^2$ is arbitrary as far as $\alpha_s(M^2)$ is small.
For $M^2$ about $(GeV)^2$ the numerical contribution
of the $\rho$-meson to the left-hand side
of (\ref{rho}) is very substantial while the right-hand side
is calculable in terms of quarks. Thus, one may say that
Eq (\ref{rho}) represents a kind of refined Sakurai duality.
Refined, in the sense that the weight function with which
one integrates the cross section is determined from the first
principles of QCD.

The weakest point was that it was not known at which $n$ (see Eq.
(\ref{psi})) or $M^2$ (see Eq.~(\ref{rho})) we should stop
applying the sum rules based on the asymptotic freedom.
Say, the number in the right-hand side of Eq.~(\ref{number})
contains a hidden dependence on the value of $n$ for which we choose
the resonance to dominate the sum rules.
For further progress, we needed
the mechanism of the AF breaking.\\

\section*{ Gluon Condensate}

The question ``who stops the Asymptotic Freedom ?''
occupied us through the whole summer of 1978 and further into the fall.
At first sight, the answer was almost trivial: the growth
of the coupling at lower momenta. It would be a common answer.
However, gradually a feeling developed that the things are not so simple.
It was difficult, however, to formalize this feeling.
Still, with time some paradoxes crystallized themselves.
For example, it was quite a common
theoretical guess that the splitting between
the vector mesons does not depend on flavor, say:
\be
m(\rho')-m(\rho)~\approx~m(\psi')-m(J/\psi)\label{simple},
\ee
which is true experimentally.
This simple-looking observation was, however, a serious challenge to the
wisdom that it is the growth
of the effective coupling (\ref{coupling})
that stops the AF at moderate mass scales.
Indeed, if expressed in terms of an invariant quantity, $s$,
Eq. (\ref{simple}) implies that the $J/\psi$ is dual to a much
larger interval of $s$ than the $\rho$
because the c-quark is heavy. In other words, the AF is violated
in the $\rho$ channel much later than in the charmonium channel,
if we start from very large $Q^2$ downwards.
A direct numerical analysis of the sum rules confirmed
this expectation. However, the coupling should run as a function
of an invariant quantity and is flavor blind.

The crucial step was to explore the possibility 
that it is the soft
nonperturbative vacuum fields that are responsible for
the stopping the asymptotic freedom.
At first sight, even having this idea would not help much
since very little is known on the precise nature of these
nonperturbative fields.
In particular, the instanton calculus was known to be very sensitive to
details of an infrared cut off \cite{gross1}.
The way out of this difficulty was not to try to calculate the
nonperturbative fields but describe
them instead phenomenologically in terms of a few parameters.

One can understand the trick by considering the same Feynman
graphs for the $\Pi_j(Q^2)$. Namely, turn to the graph with
a gluon exchange.
To ensure that it is determined by short distances
we assumed that the momentum brought by the current is large
and space-like, $Q^2\gg \La^2$. Let us consider now in more detail,
how this momentum is
transported along the
quark and gluon lines. The typical case, favored by the phase space,
is when all momenta are of order $Q$.
However, there is also a possibility that the large momentum is carried by
the quark lines alone while the gluon is soft, $k^2\ll Q^2$.
Then all the points along the quark lines are actually
very close to each other in space-time,
$\Delta x\sim 1/Q$, while the gluon line travels
far away on this scale.

Thus, under the circumstances the graph can be depicted as in
Fig. 4.
\begin{figure}[h]
  \begin{center}
    \leavevmode
    \epsfig{figure=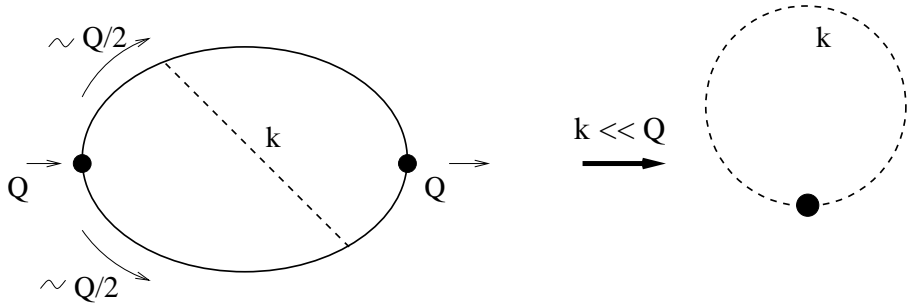,width=12cm}
    \caption{Space-time picture corresponding to the Wilson operator product expansion. The gluon, which is represented by a dashed line, is much softer
than the quarks.}
    \label{fig:fig4}
  \end{center}
\end{figure}
Moreover, there is no reason any longer to use the
perturbative expression for the gluon line since it is soft and
modified strongly by the confinement.
The quark lines propagating short distances become a receiver of
long wave gluon fields in the vacuum.
The receiver is well understood because of the asymptotic freedom.
The intensity of the gluon fields is measured. It is characterized
by the so called gluon condensate:
\be
\langle 0|\alpha_s\left( (\vec{H}^a)^2 -(\vec{E}^a)^2\right) |0\rangle
~\neq~0\label{gc}\ee
where $\vec{H}^a, \vec{E}^a$ are color magnetic and electric fields.

What is described here in words, has an adequate formulation
in terms of the Wilson operator expansion \cite{wilson}, which
allows for a systematic and straightforward calculation
of the contribution of the gluon condensate (\ref{gc})
to various correlation functions.
With account of the gluon condensate the sum rules become:
\begin{eqnarray}
&&\int R_j(s)\exp(-s/M^2)ds~\approx~(Parton~model)\nonumber\\
&& \times\left(
1+a_j{\alpha_s(M^2)\over \pi}+
c_j{\langle 0|\alpha_s(G^a_{\mu\nu})^2|0\rangle \over M^4}+...\right)
\label{qcd}
\end{eqnarray}
where the coefficients $a_j,c_j$ depend on the channel
and the ellipses stand for
higher order perturbative corrections as well as for power corrections
of higher order in $M^{-2}$.
Eqs. (\ref{qcd}) are the QCD sum rules \cite{svz}.
The first check was to see whether the gluon condensate
explains the difference in duality intervals in terms of $s$ in the
$\rho$- and $J/\psi$- channels (see above). It did explain
this difference immediately and since that moment on we did
not doubt that the gluon condensate is a real thing.
It is worth mentioning that in the world of hadrons made of
light quarks, the quark condensate, $\langle 0|\bar{q}q|0\rangle$,
is the same important as the gluon condensate \cite{svz}.
But, essentially, this is the only parameter which should be added.

The QCD sum rules turned to be a very
straightforward and successful
tool for orientation in the hadronic world
(for further references see, e.g.,
\cite{shifman}).  In a simple construct, there were unified
the perturbative physics of short
distances encoded now in the coefficients $c_j$
and the physics of the soft fields encoded
in the quark and gluon condensates.
On the theoretical side, the road was open
to introduce and treat consistently power corrections
via the Wilson operator product
expansion (earlier, the studies of the OPE 
\cite{pope} had been confined to the perturbation theory).
\section*{ Generalizations} 

We are jumping now over more than 15 years and discuss very briefly
a remarkably simple technique which allows to
consider power corrections to various observables
directly in Minkowski space.
The technique is based on introduction
of a (fictitious) gluon mass $\lambda, \lambda\to 0$
and tracing terms non-analytical in $\lambda^2$.
The idea is similar in fact to that underlying introduction of the
gluon condensate.

Indeed, let us consider again the gluon condensate but this time
in one-loop approximation and in case of a finite gluon mass.
Obviously enough, it diverges wildly in the ultraviolet.
Let us define the gluon condensate,
however, as the non-analytical in
$\lambda^2$ part of the perturbative answer \cite{chetyrkin}:
\be
\langle 0|\alpha_s (G^a_{\mu\nu})^2|0\rangle~=
{3\alpha_s\over \pi^2}\int_0^{\infty} {k^4dk^2\over (k^2+\lambda^2)}
\equiv -{3\alpha_s\over \pi^2} \lambda^4\ln\lambda^2,\label{definition}
\ee
Moreover, to finally
get rid of the gluon mass (which is not a pleasant sight for
a theorist's eye) we make a replacement:
\be
\alpha_s\lambda^4\ln\lambda^2~\rightarrow~c_4\La^4\ee
where $c_4$ is an unknown coefficient \footnote{
The technique with $\lambda\neq 0$ is in fact close to
the infrared renormalons \cite{thooft2}. In the context of the gluon
condensate, the infrared renormalons were
considered first in Ref. \cite{david}.}.
The central point is that if we evaluate various correlation functions
(\ref{correlator}) and isolate the terms $\lambda^4\ln\lambda^2$
we would reproduce the sum rules (\ref{qcd}). Indeed, as we explained
above the gluon condensate in the sum rules (\ref{qcd}) parameterizes
the infrared sensitive part of the Feynman graph associated with
a soft gluon line. Picking up terms non-analytical in $\lambda^2$
is the same good for this purpose since the non-analyticity in
$\lambda^2$ can obviously arise only from soft gluons, 
$k\sim \lambda$.

So far we have not got any new result, though.
However, the advantage of introducing $\lambda\neq 0$ is that
calculations can be performed now in Minkowski space as well
and apply for this reason to a much wider class of observables
than the OPE underlying the QCD sum rules (\ref{qcd})
\footnote{There is a price to pay, however. Infrared renormalons if
applied directly in the Minkowski space do not allow
for model independent relations between power
corrections to different observables.
The reason is that the coupling $\alpha_s$
refers now to infrared region and is, therefore, of order unit.
As a result, all orders in $\alpha_s$ are equally important,
see, e.g., \cite{vz1}.}.
The link to QCD is again through the bald replacement of the
non-analytical in $\lambda^2$ terms by corresponding powers of $\La$:
\begin{eqnarray}
\alpha_s\sqrt{\lambda^2}&\rightarrow &c_1\La \nonumber\\
\alpha_s\lambda^2\ln\lambda^2&\rightarrow & c_2\La^2~\ldots
\end{eqnarray}
where $c_{1,2}$ are some coefficients treated as phenomenological
parameters.
The gluon condensate appears now merely as one of
the terms in this sequence.

The phenomenology based on such rules turns successful
(see, e.g., \cite{review} and references therein).
The problem is not so much a lack of success but rather
too much of overlap \cite{az2}
with old-fashioned hadronization models, 
like the tube model. 

The success of this, most naive approach reveals that
at least in the cases when the technique applies the
nonperturbative effects reduce to a simple amplification
of infrared sensitive contributions to the Feynman
graphs.

\section*{ Limitations}

This simple picture is not universally true, however.
We knew it since the same year 1979 when the QCD sum rules
were formulated. The original papers
on the sum rules were 300 typewritten pages long. And many
times we were asked, how could we write such a long treatise.
It might have been not simple indeed (keeping in mind, for example,
that we made typewriting ourselves and only during weekends
when we could find a free typewriter). However, the right
question would be, I think, why we did not write the papers
much longer. Indeed, if you overstep a reasonable length of, say,
30 pages and go after 300 pages, then a curious person should have
asked, it seems,
why the papers are not, say, 3000 pages long. In fact, there was a well
defined answer to this never-asked question:
we found a first channel where the sum rules seemed not to work.
By some irony, we failed to explain appearance of a large mass scale
related to the penguins graphs, see \cite{arkady}. 
Thus we decided to make a pause to write up the cheerful part of the
story and contemplate longer about the emerging difficulties.

The difficulties did not disappear, however, by themselves but
rather deepened when we, together with Victor Novikov, came back
to the problem.
Certainly, there exist channels where the infrared sensitive corrections
described
above cannot be the whole story.
For example, we were able to show \cite{novikov}
that the leading power correction in the $0^+$-gluonium channel
is given by:
\beqn
&& \int ImG(s)\exp{(-s/M^2)}ds/s~\approx~
 \nonumber\\
&&\!\!\!\!\!\!\approx G(M^2)_{\rm parton ~model}
\left[1 +{\langle 0|\alpha_s(G^a_{\mu\nu})^2|0\rangle\over M^4}
\left(-{2\pi^2\over\alpha_s(M^2)}+{16\pi^2\over b_0\alpha_s^2(M^2)}
\right)\right]\!\!.\label{stranger}
\eeqn
where
\be
G(Q^2)\equiv i\int\! d^4x\,\exp(iqx)
\langle 0|T\{ (G^a_{\mu\nu}(x))^2,(G^a_{\mu\nu}(0))^2\}|0\rangle.
\ee
The power correction proportional to the gluon condensate
originates here from two sources. First, there is a standard OPE
correction (see (\ref{qcd})) and, second, the one evaluated
via a low energy theorem specific
for this particular channel. The correction which
is not caught by the standard OPE is about (20-30) times larger!

Thus, if we characterize the scale where the asymptotic freedom 
gets violated by the power correction by the value $M^2_{crit}$ where
these corrections become, say, 10\%, then $M^2_{crit}$ differs 
drastically in various channels:
\be
(M^2_{crit})_{\rho-meson}~\approx~0.6~GeV^2,~~~
(M^2_{crit})_{0^+~gluonium}~\approx~15~GeV^2.
\ee
Thus, the proof of the low-energy theorem brought a proof of existence
of qualitatively different scales in the hadron physics \cite{novikov}.

What made the search for the ``exceptional'' channels
where the OPE fails to identify the leading correction so difficult
was lack of any systematic way to evaluate the power corrections 
beyond the same OPE.
For example, the huge power correction in (\ref{stranger}) looked a 
stranger
since a single power correction cannot match a resonance but we
were unaware of the source of (hypothetical) other corrections
of the same mass scale.
Nevertheless, through a meticulous analysis
\cite{novikov}
we were able
to find hints
that there exists a hierarchy of the strengths
of the extra power corrections in the ``exceptional''
channels. Moreover, this hierarchy could be explained in terms
of the transitions of the corresponding currents directly to instantons,
Fig. 5 (see also \cite{ioffe}).
\begin{figure}[h]
  \begin{center}
    \leavevmode
    \epsfig{figure=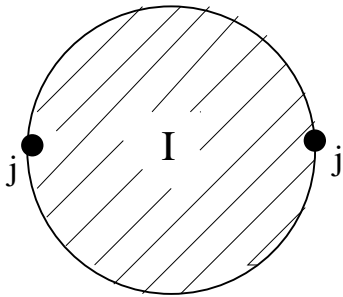,width=4.5cm}
    \caption{Direct instantons. One substitutes instanton fields, both
bosonic and fermionic
into the currents and integrates over the instanton sizes.}
    \label{fig:fig5}
  \end{center}
\end{figure}
Although qualitatively
the new extended picture worked well we left the field disappointed
by the lack of a machinery to
evaluate the new effects.
Later, a model of instanton liquid
was developed that allowed for a much more quantitative treatment
of the instanton effects
(for a review and further references see \cite{shuryak}).
The model appears to be successful phenomenologically.

\section*{ Elusive Effects of Confinement. Short Strings?}

Looking backward, it still remains a mystery
whether any specific confinement effects are revealed
through the power corrections. Indeed, consider the vacuum of pure
gluodynamics. It is known from lattice measurements
that external heavy quarks are confined by this medium
(for a review see, e.g., \cite{bali}).

On the other hand, the effects included into the sum rules so far
do not seem to encode the confinement.
Indeed, the perturbative QCD
resembles ordinary bremsstrahlung in QED.
The gluon condensate, as well as other newly found power corrections
can be detected by introducing a fictitious gluon mass which
is not sensitive to the non Abelian nature of gluons at all.
Finally, instantons are known not to ensure the confinement either
\cite{shuryak1}.

In an attempt to find power corrections related more
directly to the physics of confinement one can turn \cite{gubarev} to
the Abelian Higgs model (AHM)
which underlies the dual superconductor model
of the confinement \cite{nambu}.
If one introduces a pair of
external magnetic charges into the vacuum of the AHM model
in the Higgs phase
then the potential grows at large distances, see Eq. (\ref{lingrowth}).
The scale of distances is set up by the inverse masses
of the vector and scalar fields, $m^{-1}_{V,S}$. This growth
of the potential is due to the Abrikosov-Nielsen-Olesen
strings \cite{ano}.
The relevance to non-Abelian gauge theories is most transparent in the
so called $U(1)$ projection which treats the diagonal gluon fileds like
photons \cite{thooft}. There is ample evidence in favor of this picture
on the lattice \cite{chernodub}.

Consider now {\it short} distances, $r\ll m^{-1}_{V.S}$.
Then the Coulomb like interaction dominates. However, there
is a stringy correction to the potential at any small
distances \cite{gubarev}:
\be
\lim_{r\rightarrow 0}V(r)~=~-{Q_M^2\over 4\pi r} +\sigma_0r\label{sd}
\ee
where $Q_M$ is the magnetic charge. The ANO string is a bulky object on this
scale and is not responsible for the linear correction. Instead,
the stringy potential at short distances is due to infinitely thin
topological strings which connect the magnetic charges and which
are defined through vanishing of the scalar field along the string.
Thus, it turns out that at least in this model
the confined charges learn about confinement already at small distances
because of the short strings which are seeds for future
confining ANO strings. Amusingly enough,
it demonstrates that at short distances
a dimension two quantity is not necessarily the gluon mass squared
but could be a string tension as well. The notion of such a 
string might be
generalized to the QCD \cite{gubarev}.

The linear potential at short distances (\ref{sd})
corresponds in the momentum space to a $1/Q^2$ correction \cite{az1}.
Moreover, it can be imitated by a {\it short-distance} gluon mass.
To reproduce the positive string tension $\sigma_0)$ at short distances,
it should be however a tachyonic mass \cite{cnz}!
This, openly heuristic assumption allows to
extend the phenomenology of the $1/Q^2$ corrections.
In particular, there arises a $1/M^2$
term missing from the standard sum rules (\ref{qcd}):
\begin{eqnarray}
&&\int R_j(s)\exp(-s/M^2)ds~\approx~(Parton~model)\nonumber\\ 
&& \times\left(
1+a_j{\alpha_s(M^2)\over \pi}+
{b_j\over M^2}+
c_j{\langle 0|\alpha_s(G^a_{\mu\nu})^2|0\rangle \over M^4}+...\right)
\label{qcd1}
\end{eqnarray}
where the coefficients $b_j$ are now calculable in terms
of the tachyonic gluon mass. The modified sum rules turn to be successful
phenomenologically provided that $\lambda^2\approx-0.4GeV^2$ \cite{cnz}.
In particular, in the $O^+$-gluonium channel
a new correction arises which matches the large correction
in Eq. (\ref{stranger}) which so far was hanging without any support
from other known power-like terms.
In some channels, the new terms may compete
with the direct instanton contributions.
Further checks are necessary, however, before one can be certain
about the existence of the novel $1/Q^2$ corrections associated with
short distances\footnote{Unconventional $1/Q^2$ corrections were
introduced also within other frameworks,
such as ultraviolet renormalons, Nambu-Jona-Lasinio model, or
modified effective coupling, see, e.g., Ref. \cite{grunberg}
and references therein. The tachyonic gluon mass can be considered
as a particular prescription to fix such corrections in terms of
a single parameter.}. 

\section*{Conclusions.}

The QCD sum rules even now seem to provide a standard framework
to

(i) Analyze infrared sensitive power corrections to various 
correlation functions
and get oriented in properties of hadrons with various quantum numbers,

(ii) To look for further contributions which go beyond the
quark and gluon condensates.

The nature may turn to be generous as far as power corrections are concerned.
I am borrowing this term from a talk on dark matter. Indeed, first
people assumed that there should be a single dominant source
of the dark matter, and now it appears distributed among various
equally important components.
Similar picture may be true for the sum rules.
Indeed, very first idea would be that theoretically the 
correlation functions $\Pi_j(Q^2)$ could be found perturbatively at large $Q^2$
and
the growth of the effective coupling at smaller $Q^2$ would signal
the breaking of the asymptotic freedom. Then the picture got more involved
and the effect of soft non-perturbative fields was included
in terms of the quark and gluon condensates.
It appears not suffice to explain the peculiarities of all the channels
and the effect of direct instantons was invoked.
As the latest development, hypothetical $1/Q^2$ corrections
associated with short strings are established within
the Abelian Higgs model which is thought to mimic the QCD confinement.

Thus it appears now that all three ``ingredients'' of QCD
mentioned in the introduction have already found their way
into the sum rules. And each time there are claims of
some qualitative effects getting explained.
It might be not the end of the story.

\section*{ Acknowledgments}

On the occasion of receiving the 1999 Sakurai
Prize, it was
a pleasure for me to recollect many years of collaboration with
Misha Shifman and Arkady Vainshtein which I am acknowledging with great
gratitude. Moreover, I think that our efforts were to a great extent
simply a part
of activity of the theory groups of ITEP (Moscow) and Budker Institute 
(Novosibirsk)
and I am thankful to our colleagues for the friendly
and creative atmosphere. Collaborations with V.A. Novikov,
L.B. Okun, and M.B. Voloshin are especially acknowledged.

\end{document}